\documentclass{cjaa}                   
\usepackage{graphicx}                   
\input{epsf.sty}                        
\input{psfig.sty}                       

\begin{document}

   \title{The Current Flows in Pulsar Magnetospheres
}

   \volnopage{Vol.0 (200x) No.0, 000--000}      
   \setcounter{page}{1}          

   \author{Ren-Xin Xu, Xiao-Hong Cui, Guo-Jun Qiao
      }
   \offprints{Ren-Xin Xu}                   

   \institute{Astronomy Department, School of Physics, Peking University, Beijing 100871, China\\
             \email{r.x.xu@pku.edu.cn}
          }

   \date{Received~~~~~~~~~~~~~~~~~; ~~accepted}

   \abstract{
The global structure of the current flows in pulsar magnetospheres
is investigated, with rough calculations of the elements in the
magnetospheric circuit.
It is emphasized that the potential of critical field lines is the
same as that of interstellar medium, and that the pulsars whose
rotation axes and magnetic dipole axes are parallel should be
positively charged, in order to close the pulsar's current flows.
The statistical relation between the radio luminosity and pulsar's
electric charge (or the spindown power) may hint that the
millisecond pulsars could be low-mass bare strange stars.
   }

   \authorrunning{Xu, Cui \& Qiao}            
   \titlerunning{The Current Flows in Pulsar Magnetospheres}  

   \maketitle
%
\section{Introduction}

Since Hewish et al. (1968) discovered the first radio pulsar, more
and more magnetospheric models for pulsars have been proposed to
explain their observed phenomena. The vacuum inner gap model, on
the one hand, suggested first by Ruderman \& Sutherland (1975,
hereafter RS75) depends on enough binding energy of charged
particles on the pulsar surface. The space charge-limited flow
model (e.g., Arons \& Scharlemenn 1979; Harding \& Muslimov 1998),
on the other hand, is without any binding energy. An outer gap
near light cylinder has also been proposed (e.g., Cheng, Ho, \&
Ruderman 1986), the existence of which may also reflect strong
binding of particles on pulsar surface (Xu 2003a).
However, as the observational data in radio, optical, X-ray, and
$\gamma$-ray bands accumulates, there are still a great number of
puzzles to be solved (e.g., Melrose 2004).
Nevertheless, it is obvious that the models depend on the nature
of pulsar surfaces, and one may conclude on the interior structure
of pulsars (e.g., either normal neutron stars or bare strange
stars, see, e.g., Xu 2003b), which is almost impossible via
calculations in supranuclear physics, via investigating pulsar
emission models.

The study of pulsar magnetosphere is essential for us to
understand various radiative processes, and thus observed emission
in different bands.
Goldreich and Julian (1969) argued that a pulsar must have a {\em
magnetosphere} with charge-separated plasma and demonstrated that
a steady current would appear if charges can flow freely along the
magnetic field lines from the pulsar surface.
Sturrock (1971) pointed that such a steady flow is impossible due
to pair formation, and suggested that a simple electric circuit of
the pulsar magnetosphere would be a discharge tube connected with
an electromotive source.
RS75 proposed that sparking process takes place in a
charge-depletion gap just above the pulsar surface. The sparking
points drift due to ${\bf E}\times {\bf B}$, which can naturally
explain drifting subpulse phenomena observed in radio band.
Based on the assumption that the magnetosphere has a global
current loop which starts from the star, runs through the outer
gap, the wind and the inner gap, and returns to the star, Shibata
(1991) proposed a circuit including an electromotive source
connected in series with two accelerators (the inner and outer
gaps) and the wind.
Providing a fully general relativistic description, Kim et al.
(2005) studied the pulsar magnetosphere and found that the
direction of poloidal current in neutron star magnetosphere is the
same as that in black hole.

In this work, assuming that the critical magnetic-field lines are
at the same electric potential\footnote{%
We choose the potential of the interstellar medium to be zero,
$\phi_{\rm ISM}=0$, in this paper.
} %
as the interstellar medium (ISM, Goldreich \& Julian 1969), we
propose a circuit model for pulsar magnetosphere.
The corresponding relation of the elements between magnetosphere
and circuit is as follows.
(1) The total inner radiation of the pulsar corresponds to an
electromotive power parallel connecting with a capacitor. Note
that the resistance of power is negligible due to perfect
conductivity of the star.
(2) The inner gaps which include the inner core gap (ICG) and the
inner annular gap (IAG) (Qiao et al. 2004a, 2004b) correspond to a
parallel connection of a resistor and a capacitor. When a spark
takes place in gap, it can be represented by a resistor; if there
is no spark, the voltage on the gap is so high that it can be
described by a capacitor.
(3) The outer gap and the pair-plasma wind correspond to a series
connection about a inductor and a parallel connection of a
resistor and a capacitor.
Considering a {\em parallel} pulsar whose rotation axis is
parallel to the magnetic axis, we find that the pulsar should be
positively charged on the surface for the necessarian of a close
circuit. The total electric field along field lines,
$E_\parallel$, in the magnetosphere is then composed of two
components: that due to charge-departure from the Goldreich-Julian
density, and that due to the charges of pulsar.

This paper is arranged as follows. The model is introduced in \S
2. The total charges of pulsar are estimated in \S 3. In \S 4, we
discuss about the charges of low-mass strange stars and show
evidence for low-mass millisecond pulsars by observational data.
Conclusions are presented in \S 5.

\section{The model}

As shown in Fig.1, the foot points of line {\em a, b}, and
rotation axis on the star surface are assigned as {\em A, B} and
{\em P}. Point {\em P} is also the magnetic pole of star. We
assume that the potential of the critical field line {\em b}
equals to that of ISM at infinity, $\phi_B=0$; otherwise, a close
electric current in the two regions (i.e., {\em Region I}: with
boundary lines labelled ``a'', and {\em Region II}: that between
lines ``a'' and ``b'') of open field lines is
impossible\footnote{%
For {\em parallel} pulsars, electric current flows outward in the
two regions if one sets the potential of the polar line to equal
to that of ISM, $\phi_P=0$, since all the potentials of open-field
lines, except the polar line, are greater than zero. Also one can
see that the current flows inward in those two regions if one
chooses the potential of the last open-field lines to be zero,
$\phi_A=0$. Current flows can {\em not} be closed in both these
cases.}.
Then for a {\em parallel} pulsar, the potential $\phi_I<0$ (within
{\em Region I}) and $\phi_{II}>0$ (within {\em Region II}) in the
regime of $\phi_B=0$. Therefore, the negatively charged particles
should flow out along the open magnetic field lines within {\em
Region I} from star, but positively charged particles flow out
through {\em Region II}.

\begin{figure}
  \centering
    \includegraphics[width=8cm]{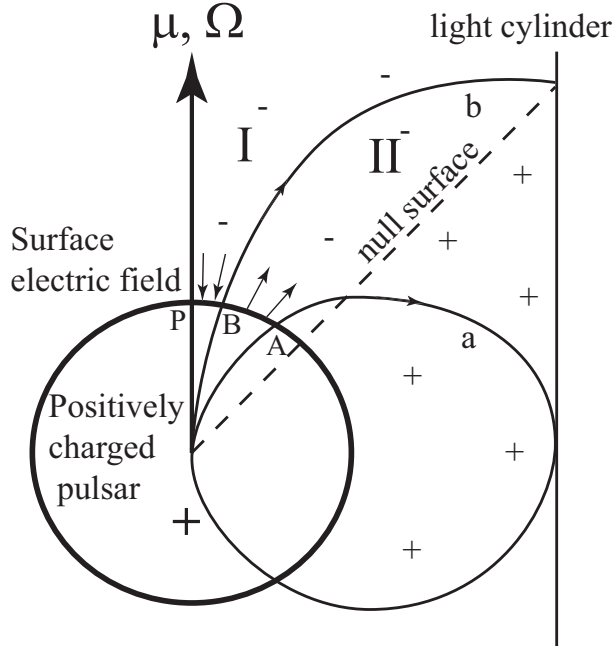}
    \caption{
Sketch of the magnetic field line, electric field line, and
magnetospheric charge distribution for a {\em parallel} pulsar.
Line {\em a} is the last open magnetic field line. Line {\em b} is
the critical field line. {\em Region I}: with a boundary of lines
labelled ``b'', and {\em Region II}: between lines ``a'' and
``b''. If the potential of line {\em b} equals to that of
interstellar medium, then the potential within {\em Region II} is
positive but negative within {\em Region I}.}
\end{figure}

The inner gaps (including ICG and IAG in this paper) and an outer
gap may work in a magnetosphere\footnote{%
The inner annular gap and the outer gap might not exist
simultaneously.
}. %
There is no current in the circuit until a spark takes place in
the inner gaps, so these gaps could be simulated by parallel
connected a resistor and a capacitor, which connect with other
parts in series (as shown in Fig.2).
\begin{figure}
  \centering
    \includegraphics[width=8cm]{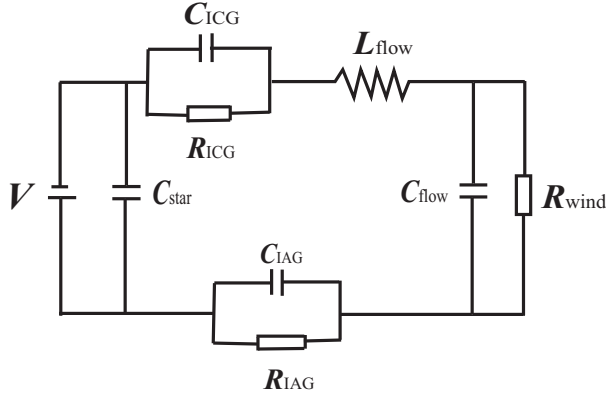}
    \caption{An equivalent circuit of pulsar magnetosphere.
    An electromotive source connected with other parts in series
    corresponding to the ICG, IAG, the outer gap and the pair-plasma wind. The current
    is determined by the power voltage {\em V}. ``{\em R}'' stands for plasma resistance effects if a spark happens,
    mainly the losses of relativistic particles. ``{\em C}'' represents the gap capacitance
    effects if there is a potential drop on the gap. ``{\em L}'' describes the electromagnetic characteristic of flow.}
\end{figure}

In Fig.2, the power of star could be equivalently modelled by
voltage $V$ and capacitance $C_{\rm star}$.
The star is magnetized and possesses an interior electric field,
$\vec{E}$, which satisfies
\begin{equation}
\vec{E}+\frac{\vec{\Omega}\times \vec{r}}{c} \times \vec{B}=0,
\end{equation}
where $\Omega$ is the angular velocity of star rotating around the
dipole rotation axis, which relates with rotating period of star
{\em P} by $P=2\pi/\Omega$. The magnetosphere of a rotating
isolated-pulsar is thus generally concluded to be powered by an
electric source with certain potential drop between the polar
angle $\theta_{\rm B}$ and $\theta$ of
\begin{equation}
\phi=\frac{R^{2}\Omega B}{2c}(\sin^{2}\theta-\sin^{2}\theta_{\rm
B})\approx3\times
10^{16}R_6^2B_{12}P^{-1}(\sin^{2}\theta-\sin^{2}\theta_{\rm
B})~~{\rm volts},
\label{V}%
\end{equation}
if the pulsar is assumed to be magnetized homogenously, where
$R_6=R/(10^6$ cm), $B_{12}=B/(10^{12}$ G). According to the
equation of dipolar field line, $r=r_{\rm d}\sin^{2}\theta$
($r_{\rm d}$ is the maximum polar radius), and the polar angle of
null surface $\theta_{\rm n}=\cos^{-1}(\pm 1/\sqrt{3})$, one can
obtain $\sin^{2} \theta_{\rm B}=(2/3)^{3/2}R/R_{\rm L}$ , where
the radius of light cylinder $R_{\rm L}\equiv c/\Omega=cP/(2\pi)$.

The capacitance of the star with a radius $R$ is, in {\em cgs}
units,
\begin{equation}
C_{\rm star} = 9\times 10^{-21} R~~{\rm farads.}%
\label{Cstar}
\end{equation}
The electric flow process in a magnetosphere can be equivalently
mimicked by an antenna cable with capacitance and inductance. They
could be equivalently described by the inductance $L_{\rm flow}$
and $C_{\rm flow}$, the values of which can be estimated as
\begin{equation}
C_{\rm flow}\simeq 2\pi\varepsilon_0l(\ln {r_2\over r_1})^{-1}%
\end{equation}
and
\begin{equation}
L_{\rm flow}\simeq {\mu_0\over 2\pi}l\ln {r_2\over r_1}%
\end{equation}
respectively; where $l$, $r_1$, and $r_2$ are the length, inner
radius, and outer radius of the cable, respectively. The passage
length of the electric current $l\sim \sqrt{3/2}r_{\rm
L}=5.8\times 10^7P$ m. Near the light cylinder, the ratio
$r_2/r_1\sim (r_{\rm L}+r_{\rm L}/\sqrt{2})/r_{\rm L}=1.7$; but on
the stellar surface, the ratio $r_2/r_1=(3/2)^{3/4}=1.4$. Both
them are not dependent on $P$, $B$, and other parameters, and we
choose thus simply $r_2/r_1=1.5$ in this paper.
One therefore comes to
\begin{equation}
C_{\rm flow}\sim 8.0\times 10^{-3}P~~{\rm farads},%
\label{Cflow}%
\end{equation}
and
\begin{equation}
L_{\rm flow}\sim 4.7P~~{\rm henries}.%
\label{Lflow}%
\end{equation}

In case enough binding energy of charged particles on the stellar
surface, an RS-type (RS75) vacuum gap should exist near polar cap,
which can be the equivalent of a capacitor of two parallel slabs,
with
\begin{equation}
C_{\rm RS}=\pi \varepsilon_0 r_{\rm p}^2/h=5.8\times
 10^{-8}R_6^3Ph_3^{-1}~~{\rm farads},%
\label{Cflow}%
\end{equation}
where $r_{\rm p}=R\sin\theta_{\rm p}=1.45\times 10^4
R_6^{3/2}P^{1/2}$ cm is the radius of polar cap, the gap height
{\em h} is a model dependent parameter, $h_3=h/(10^{3}$ cm).

For the curvature-radiation-induced and the resonant
inverse-Compton-scattering-induced cascade models, the gap heights
{\em h} could be (e.g., Zhang, Harding, \& Muslimov 2000)
\begin{equation}
h_{\rm cr}=54\rho_6^{2/7}B_{12}^{-4/7}P^{3/7}~~{\rm m},%
\label{hcr}%
\end{equation}
and
\begin{equation}
h_{\rm ics}=279\rho_6^{4/7}B_{12}^{-11/7}P^{1/7}~~{\rm m},%
\label{hcr}%
\end{equation}
respectively, where $\rho$ is the radius of curvature of field
line, $\rho_6=\rho/(10^6$ cm).

If the field lines which cross the light cylinder can only not
co-rotate, the active region of the star is then the polar cap,
from polar angle $0$ to $\theta_{\rm p}\simeq \sqrt{2\pi
R/(cP)}=1.45\times 10^{-2}(R_6/P)^{-1/2}$. In this case, the
potential difference of the electrical source is therefore
\begin{equation}
V_{\rm cap}=-6.58\times 10^{12}R_6^3B_{12}P^{-2}~~{\rm volts}.%
\label{Vcap}%
\end{equation}

Since the dominate source of rotation energy dissipation is
through $R_{\rm wind}$, we can estimate $R_{\rm wind}$ as
\begin{equation}
R_{\rm wind}\simeq {V_{\rm cap}^2\over {\dot E}_{\rm rot}}=
11M_1^{-1}R_6^4B_{12}^2{\dot P}_{15}^{-1}P^{-1}~~{\rm ohms},%
\label{Rwind}%
\end{equation}
where the rotation loss rate ${\dot E}_{\rm rot}=-4\pi^2I{\dot
P}/P^3$, $I\sim 10^{45}M_1R_6^2$ g$\cdot$cm$^2$ is the moment of
inertia for a neutron star with mass $\sim M_1 M_\odot$ and radius
$\sim R_6 \times 10$ km, the period derivative ${\dot
P}_{15}=|{\dot P}|/10^{-15}$. Compared with $R_{\rm wind}$, the
stellar resistance $R_{\rm star}$ is negligible due to the perfect
conductivity of the star.

The potential drop of outer gap, where no spark happens,
corresponds to a resistor, which is presumed to be combined with
the wind dissipation as a total one $R_{\rm wind}$ in Fig.2. In
the magnetodipole radiation model, the filed $B$ and the spindown
$\dot P$ is connected by (e.g., Manchester \& Taylor 1977)
$B_{12}=3.2\times 10^7\sqrt{P{\dot P}}$, we then have
\begin{equation}
R_{\rm wind}=11M_1^{-1}R_6^4~~{\rm ohms}.
\end{equation}

For the potential drops of inner gaps (ICG and IAG), when the
spark provides necessary charges to close the pulsar circuit, they
should correspond to resistors ($R_{\rm IAG}$ and $R_{\rm ICG}$,
shown in Fig.2). When there is no sparks, the gap grows and they
can be described by capacitors ($C_{\rm IAG}$ and $C_{\rm ICG}$,
shown in Fig.2).

Let's analyze the circuit in Fig.2. Although the electric power
has fixed potential supply, the current is changing due to the
inner gap sparks. In this sense, the equivalent circuit
description in this paper is not simply a DC circuit model.
Because of the erratic sparking, the resistance $R_{\rm RS}$ could
be as the sum of many sinusoidal functions of time, $R_{\rm
RS}=\Sigma_{n=0}^{\infty} R_n\sin{n\omega t}$. The electric
current between arbitrary points {\em M} and {\em N} in circuit
could also be in this form, $I_{\rm MN}=\Sigma_{n=0}^{\infty}
I_n\sin{n\omega t}$.

According to the Kirchhoff's current and voltage laws, the complex
impedances of parallel connection circuits composed of $R_{\rm
ICG}$ and $C_{\rm ICG}$, $R_{\rm IAG}$ and $C_{\rm IAG}$, $R_{\rm
wind}$ and $C_{\rm flow}$ are $z_{\rm ICG}=R_{\rm ICG}/(1+i\omega
R_{\rm ICG}C_{\rm ICG}), z_{\rm IAG}=R_{\rm IAG}/(1+i\omega R_{\rm
IAG}C_{\rm IAG}), z_{\rm wind}=R_{\rm wind}/(1+i\omega R_{\rm
wind}C_{\rm flow})+i\omega L_{\rm flow}$, respectively,
where $i=\sqrt{-1}$, $\omega$ the angular frequency of electric
current modulation. Defining $z'\equiv z_{\rm ICG}+z_{\rm
IAG}+z_{\rm wind}$, one obtains the total complex impedance to be
\begin{equation}
z_{\rm total}=\frac{z'}{1+i\omega z'C_{\rm star}}~~{\rm ohms}.
\end{equation}

Let $\omega=\omega_0$ when $|z_{\rm total}|$ is the smallest
value. In this case, the potential drop between inner vacuum gap
is the highest. It is possible that there exists an oscillation
with time scale $\omega_0^{-1}$ in the circuit.
We expect that some of the variations of radio intensity in
different timescales could be hints of such circuit oscillations.

In case of $C_{\rm ICG}=C_{\rm IAG}=L_{\rm flow}=C_{\rm flow}=0$,
one has the total impedance $z=R_{\rm ICG}+R_{\rm IAG}+R_{\rm
wind}$. The physical meaning of this result is of the DC circuit
model (Shibata 1991), where ${\dot E}_{\rm rot}\simeq 2\pi r_{\rm
p}^2c\rho_{\rm GJ}\cdot V_{\rm cap}$, which results in the
deduction of pulsar magnetic fields $B\simeq\sqrt{c^3IP{\dot
P}/(\pi^2R^6)}$ (e.g., Manchester \& Taylor 1977), is usually
assumed (e.g., Xu \& Qiao 2001), with the Goldreich-Julian density
$\rho_{\rm GJ}$.

\section{Pulsars charged electrically?}

If charged particles distribute as the Goldreich-Julian density,
$\rho_{\rm GJ}$, they will be in balance about electrostatic
force. The equivalent ``Possion'' equation in comoving frame then
is (e.g., Beskin, Gurevich \& Istomin 1993)
\begin{equation}
\nabla\cdot \vec{E}=4\pi(\rho-\rho_{\rm GJ}).
\label{EGJ}
\end{equation}
As mentioned by RS75, the electric field, $\vec{E}_{\rm GJ}$, on
star surface due to the lack of charge density respective to the
Goldreich-Julian density (e.g., for vacuum outside the star,
$\rho=0$) is normal to star surface.
The solution of RS75 (see its appendix I.$b$) for the electric
field on the stellar surface is $E_{\rm s}=-2\Omega B h/c<0$,
which is equivalent to that of choosing the potential of field
line {\em a} to be the same one of the ISM, $\phi_A=0$. From
equation (\ref{EGJ}), one can also find generally the surface
electric field $E_{\rm GJ}<0$ in both {\em Regions I} and $II$, as
shown in Fig.3.

When the potential of the line {\em b} is zero, as mentioned
above, the vector direction of electric field  $\vec{E}_{\rm I}$
within {\em Region I} is inward  but that in {\em Region II},
$\vec{E}_{\rm II}$, is outward.
How can one understand consistently this picture? Why should this
be reasonable if one choose $\phi_B=0$, rather than $\phi_A=0$?
The answer could be that there must be positive charges on the
star surface which increases the potential of star. These charges
provide a monopole electric field $\vec{E}_{\rm mo}$, which
combined with $\vec{E}_{\rm GJ}$ to form the total electric field,
as shown $\vec{E}_{\rm I}$ and $\vec{E}_{\rm II}$ in Fig.3.

\begin{figure}
  \centering
    \includegraphics[width=10cm]{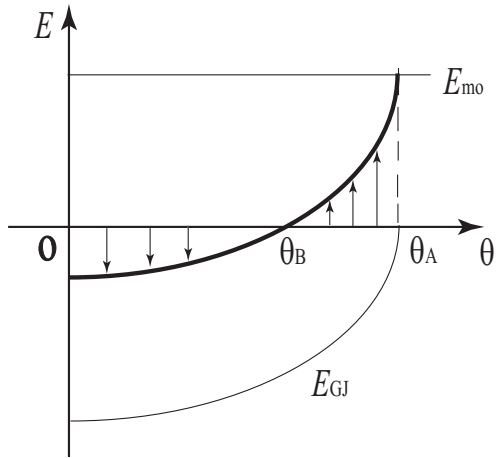}
    \caption{An illustration of the total electric field $\vec{E}$ as a function of
    polar angle $\theta$. $\vec{E}_{\rm GJ}$ is due to the charge-separation in pulsar magnetosphere, provided that
    the potential of line {\em a} is the same of the ISM. $\vec{E}_{\rm mo}$ is induced by
    the charges of star.
    The total electric field (middle solid thick curve), $\vec{E}_{\rm I}$ and $\vec{E}_{\rm II}$ within
    the corresponding regions, could be the combination of $\vec{E}_{\rm GJ}$ (lower solid thin curve) and
    $\vec{E}_{\rm mo}$ (upper solid thin straight line).}
\end{figure}

The electricity induced by the charges on the star surface is so
high that the electric field is reversed across {\em Region II}
when electric field of star self and the field caused by charged
particles are combined. Therefore, the current can flow out and
come back in magnetosphere to close the pulsar's generator
circuit. Reversely, the increased electric field supports our
assumption that the potential of line {\em b} is zero.

Provided that two conditions are satisfied (1, the star has
positive charges on the surface; 2, the potential of the line {\em
b} is zero), the potential increase of star relative to that of
ISM could then be estimated to be order of the potential drop
between {\em A} and {\em B} (from equation 2)
\begin{equation}
V_{\rm star}\sim \phi_{\rm A}-\phi_{\rm
B}=\frac{3\times10^{16}R^{2}_{6}B_{12}}{P}(\sin^{2}\theta_{\rm
A}-\sin^{2}\theta_{\rm B})
\approx \frac{3\times10^{12}R^{3}_{6}B_{12}}{P^{2}}~~~{\rm volts}.
\label{Vstar}
\end{equation}
From equations (\ref{Cstar}) and (\ref{Vstar}), the total charges
on stellar surface could then be
\begin{equation}
Q=V_{\rm star}\times C_{\rm star}\approx
\frac{3\times10^{-3}R^{4}_{6}B_{12}}{P^{2}}~~~{\rm coulombs}.%
\label{Q}
\end{equation}
The collapse of an evolved massive star should form temporally a
rotating magnetized compact star, and then become a black hole.
The black hole could be charged too, with a quantity to be
proportional to $B/P^2$.
As magnetic field $B\sim \sqrt{P{\dot P}}$, one has then $Q\sim
\sqrt{{\dot P}/P^3}\sim {\dot E}_{\rm rot}^{1/2}$, where the
rotation energy loss rate ${\dot E}_{\rm rot}\sim \Omega{\dot
\Omega}$.
Observationally, the X-ray luminosity could be a function of the
spin-down energy loss for all rotation-powered pulsars, $L_{\rm
x}\sim {\dot E}_{\rm rot}$ (Becker \& Tr\"umper 1997); but the
$\gamma$-ray luminosity $L_\gamma\sim {\dot E}_{\rm rot}^{1/2}$
(Thompson 2003).
One can also address that energetic $\gamma$-ray luminosity is
proportional to the electric charge $Q$, rather than to ${\dot
E}_{\rm rot}$.

One thinks conventionally, according to equation (\ref{EGJ}), that
{\em no} acceleration (e.g., $E_\parallel=0$) occurs if
$\rho=\rho_{\rm GJ}$ in pulsar magnetosphere. We note, however,
that this conclusion is valid only if {\em no} solenoidal force
field appears.
In other words, $E_\parallel\neq 0$ though $\rho=\rho_{\rm GJ}$ if
one adds any solenoidal force field in the magnetosphere.
A charged pulsar contributes a solenoidal electric field, which
results in an acceleration near the pulsar (to damp as $1/r^2$).
In the close field line region, this electric field causes a
re-distribution so that $E_\parallel=0$.
In the open field line region, this field accelerates particles
(for {\em parallel} pulsars, to accelerate negative particles in
{\em Region I}, but positive particles in {\em Region II}; see
Fig.1). Certainly, extra acceleration due to $\rho\neq \rho_{\rm
GJ}$ exists too.
As demonstrated in Fig.3, a very large acceleration may exist near
the last open field lines, which could be favorable for the high
energy emission in the caustic model (Dyks \& Rudak 2003).
An outer gap may not be possible if particles can flow out freely
either from the surface (for negligible binding energy) or from
the pair-formation-front (for enough binding energy) of a charged
pulsar.

\section{Low-mass bare strange stars}

Pulsars could be bare strange stars, some of them could be of
low-mass (Xu 2005).
Due to the color self-confinement of quark matter, the density of
low-mass bare quark star is roughly homogeneous, and its mass
would be
\begin{equation}
M_{\rm QS}=\frac{4}{3}\pi
R^{3}(4\bar{B})=0.9R^{3}_{6}\bar{B}_{60}M_{\odot},%
\label{M}
\end{equation}
where the bag constant $\bar{B}=60\bar{B}_{60}$ MeV $\rm fm^{-3}$
(i.e. $1.07\times10^{14}$g $\rm cm^{-3}$). For a star with pure
dipole magnetic field and a uniformly magnetized sphere, the
magnetic moment is
\begin{equation}
\mu=\frac{1}{2}BR^{3}.
\end{equation}
If the magnetized momentum per unit mass is a constant $\mu_{\rm m
}=(10^{-4}\sim 10^{-6})\rm G \cdot\rm cm^{3}\cdot\rm g^{-1}$, the
magnetic moment then is (Xu 2005)
\begin{equation}
\mu=\mu_{\rm m }M.%
\label{mu}
\end{equation}
Combing equations (\ref{M})-(\ref{mu}), one can obtain the
magnetic field stength
\begin{equation}
B=1.8\times10^{-18}\mu_{\rm m}\bar{B}_{60}M_{\odot}.
\end{equation}

Therefore, for a low-mass bare strange star, if the values of its
period $P$, radius {\em R} (or the mass $M_{\rm QS}$), and the
polar magnetic field $B$ (or the parameter $\mu_{\rm m}$) are all
known, the total charges on the surface could be obtained by
equation (\ref{Q}).

\subsection{Evidence for low-mass millisecond pulsars?}

Pulsar's radius is assumed as a constant in above sections. In
fact, the radius should be a variable in different models (of
either normal neutron star or strange quark star). The radius of
bare strange stars could be as small as a few kilometers (even a
few meters).
Could one find any observational hints about the star radius? It
is suggested that normal pulsars could be bare strange stars with
solar masses, whereas millisecond pulsars are of low masses (Xu
2005). Can we show evidence for low-mass millisecond pulsars in
the bare strange star model for pulsars? These are investigated,
based on the observational pulsar
data\footnote{http://www.atnf.csiro.au/research/pulsar/psrcat/}.
There are 1126 pulsars with known {\em P}, $\dot{P}$ and radio
luminosity $L_{1400}$ (mJy kpc$^2$) at 1400 MHz simultaneously.
The numbers of millisecond and normal radio pulsars are 35 and
1091, respectively, if the dividing line of them is $P=15$ ms.
Here we assign the 35 millisecond radio pulsars ($P<15$ ms) as
{\em Sample I} and the 1091 normal radio pulsars ($P>15$ ms) as
{\em Sample II}.

Assuming the pulsar radius is a variable, from equation (\ref{Q}),
one can obtain pulsar's charges $Q\sim R^4 \frac{B}{P^2}\sim
R^4(\frac{\dot{P}}{P^3})^{1/2}$ since the magnetic field strength
$B\sim\sqrt{P\dot{P}}$. Defining $\zeta\equiv\frac{\dot{P}}{P^3}$,
one comes to
\begin{equation}
\log Q\sim 4\log R+\frac{1}{2}\log\zeta.%
\label{Qeta}
\end{equation}
At the same time, the rotation energy loss rate ${\dot E}_{\rm
rot}\simeq I \Omega{\dot \Omega}\sim R^5\zeta$, where rotational
inertia of star $I\simeq \frac{2}{5}MR^2\sim R^5$ (the star is
assumed to be a homogeneous rigid sphere), i.e.,
\begin{equation}
\log{\dot E}_{\rm rot}\sim 5\log R+\log \zeta.%
\label{Edoteta}
\end{equation}

The correlation between $L_{1400}$ and $\zeta$ and the normalized
statistical distribution of $L_{1400}/\zeta$ are shown in Fig.4.
For showing the evidence for low-mass millisecond pulsars by this
relation and distribution, we firstly give the best fit line by
the least square method for larger {\em Sample II}. Secondly, we
assign the slope of fitting line for the smaller {\em Sample I} to
be the same as that of {\em Sample II}. Thirdly, we study the
relation of $L\sim R$ with two assumptions, respectively, that are
the luminosity $L=L(Q)$ (i.e., $L$ is only a function of $Q$) and
$L=L({\dot E}_{\rm rot})$ (i.e., $L$ is only a function of ${\dot
E}_{\rm rot}$). Finally, through comparing the intercepts of two
fit lines, we can obtain the ratio of radius between {\em Sample
I} and {\em Sample II}.
\begin{figure}
  \centering
    \includegraphics[width=12cm]{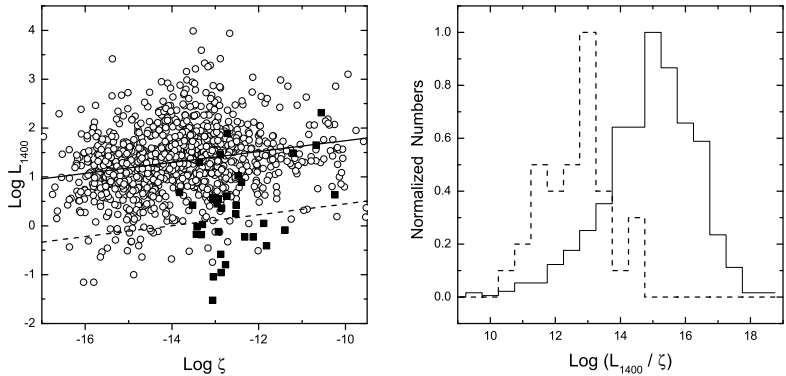}
    \caption{Left: Observed luminosity $L_{1400}$ (mJy kpc$^2$) at
    1400 MHz as a function of $\zeta$ for 1126 radio pulsars, where
$\zeta\equiv\frac{\dot{P}}{P^3}$. The solid squares represent the
millisecond radio pulsars ({\em Sample I}) with $P<15$ ms while
the empty circles are the normal ones ({\em Sample II}) with
period $P>15$ ms. The solid line is the best fit result with the
least square method for {\em Sample II}, and the dash line is the
fit for {\em Sample I} assigned the slope to be the same as that
of {\em Sample II}. Right: The normalized statistical distribution
of $L_{1400}/\zeta$ for {\em Sample I} (dash step line) and {\em
Sample II} (solid step line). The vale of $L_{1400}/\zeta$ is only
a function of pulsar radius according to
Eq.(\ref{Qeta}-\ref{L1400}).}
\end{figure}

From left panel in Fig.4, we find that the best fit line for {\em
Sample II} are
\begin{equation}
\log L_{1400}=2.86+0.11\log\zeta.%
\label{L1400}
\end{equation}
Assigning the same slope for {\em Sample I}, we find that the
intercept of fit line for this sample is 1.56.

If $L\sim Q$, from equations (22) and (24), the ratio of radii
$R_{II}$ for {\em Sample II} and $R_{I}$ for {\em Sample I} at
1400 MHz then should be $4\log\frac{R_{II}}{R_I}=2.86-1.56=1.30$.
Then one can obtain $R_{II}\approx2.11R_{I}$.
In the same way, if $L\sim {\dot E}_{\rm rot}$, from equations
(23) and (24), we can obtain $R_{II}\approx1.82R_{I}$.
From right panel in Fig.4, it is evident that the values of
$L_{1400}/\zeta$, as a function of pulsar radius, distribute with
two peaks; the higher one (i.e., larger radius) for normal pulsars
whereas the lower one (i.e., smaller radius) for millisecond
pulsars.

Summarily, the radii of {\em Sample II} are always larger than
those of {\em Sample I}, which gives an evidence that the radii of
millisecond pulsars are smaller and the masses are lower than
those of normal radio pulsars. Especially, from the right panel in
Fig.4, we can find that the normalized distributions of
$L_{1400}/\zeta$ for {\em Sample I} and {\em Sample II} are almost
clustered at about 13 and 15. This two-peak structure gives a
crude same result and thereby supports again the evidence for
low-mass millisecond pulsars.
If one thinks that normal pulsars are bare strange stars with mass
$\sim M_\odot$, millisecond pulsars could be bare strange stars
with mass $\sim (1/2)^3M_\odot$.

\section{Conclusions}

Assuming that the magnetosphere of pulsar has a global current
which starts from the star, runs through the inner core gap, the
wind, the outer gap and the inner annular gap, and returns to the
star, we study the circuit characteristics of four elements: the
electromotive source, the inner core gap, the inner annular gap,
and the outer gap and wind. It is emphasized that the potential of
critical field lines equals to that of interstellar medium. We
find, in this case, that the pulsar whose rotation axis and
magnetic dipole axis are parallel should be positively charged for
the pulsar's generator circuit to be closed. The current flows out
through the light cylinder and then flow to the stellar surface
along the open magnetic field line.

There are five independent parameters which can describe
completely the dynamics of a pulsar magnetosphere assuming a
dipole field and a uniform density of the star. They are the
radius $R$, the mass $M$($\simeq 16\pi{\bar B}R^3/3$ in case of
bare strange stars, with ${\bar B}$ the bag constant), the
magnetic strength $B$ (or magnetic moment $\mu\sim BR^3$), the
period $P$, and inclination angle $\alpha$.
Typical parameters for radio pulsars are: $R\sim 10^{6}$ cm,
$M\sim 1.4 M_{\odot}$, $P\sim (10^{-3}-1)$ s, $B\sim 10^{8-12}$ G.
No solid observational evidence shows these parameters are really
{\em typical}.
In case that pulsars are bare strange stars (probably with low
masses), some of the parameters above may not be representative,
and the independent parameters could be related each other (Xu
2005).
The statistics between the radio luminosity and pulsar's electric
charge (or the spindown power) may hint that millisecond pulsars
could be low-mass bare strange stars (with masses of a few
$0.1M_\odot$).


{\em Acknowledgments}:
The authors thank helpful discussion with the members in the
pulsar group of Peking University. This work is supported by NSFC
(10273001) and the Special Funds for Major State Basic Research
Projects of China (G2000077602).


\end{document}